\documentclass[twoside]{article}
\usepackage{Proc_NTSE}

\usepackage{graphicx}
\usepackage{isotope}
\usepackage{cite}
\DeclareRobustCommand{\abs}[1]{\lvert#1\rvert}
\newcommand{\sumprime}{\sideset{}{'}\sum}
\newcommand{\mcvec}[1]{\boldsymbol{\mathbf{\mathrm{#1}}}}
\newcommand{\uvec}[1]{\boldsymbol{\mathbf{\mathrm{\hat{#1}}}}}
\DeclareRobustCommand{\tket}[1]{\vert #1 \rangle}
\DeclareRobustCommand{\tme}[3]{\langle {#1} \vert {#2} \vert {#3} \rangle}
\DeclareRobustCommand{\toverlap}[2]{\langle #1 \vert #2 \rangle}

\newcommand{\Nmax}{{N_\text{max}}}
\newcommand{\Ntot}{{N_\text{tot}}}

\newcommand{\Nex}{{N_\text{ex}}}

\newcommand{\Trel}{{T_\text{rel}}}

\newcommand{\Ncm}[1][]{{N_\text{c.m.}^{#1}}}  
  
\newcommand{\Ncut}{{N_\text{cut}}}  

\newcommand{\wfho}{\Psi}

\newcommand{\MeV}{{\mathrm{MeV}}}
\newcommand{\fm}{{\mathrm{fm}}}

\newcommand{\hw}{{\hbar\Omega}}
\newcommand{\bint}{{b_{\text{int}}}}
\newcommand{\hwint}{{\hbar\Omega_{\text{int}}}}

\begin{document}
\thispagestyle{plain}

\begin{center}
{\Large \bf \strut
Halo nuclei with the Coulomb-Sturmian basis
\strut}\\
\vspace{10mm}
{\large \bf 
M.~A.~Caprio$^{a}$, P.~Maris$^{b}$ and J.~P.~Vary{$^b$}}
\end{center}

\noindent{\small $^a$\it Department of Physics, University of Notre Dame, Notre Dame, Indiana 46556, USA} \\
{\small $^b$\it Department of Physics and Astronomy, Iowa State University, Ames, Iowa 50011, USA}

\markboth{M.~A.~Caprio, P.~Maris and J.~P.~Vary}{Halo nuclei with the Coulomb-Sturmian basis} 

\begin{abstract}
The rapid falloff of the oscillator functions at large radius
(Gaussian asymptotics) makes them poorly suited for the description of
the asymptotic properties of the nuclear wave function, a problem
which becomes particularly acute for halo nuclei. We consider an
alternative basis for \textit{ab initio} no-core configuration
interaction (NCCI) calculations, built from Coulomb-Sturmian radial
functions, allowing for realistic (exponential) radial falloff.  NCCI
calculations are carried out for the neutron-rich
\isotope{He} isotopes, and estimates are made for the RMS radii of the
proton and neutron distributions.
\\[\baselineskip] 
\end{abstract}

\section{Introduction}

The \textit{ab initio} theoretical
description of light nuclei is based on direct solution of the nuclear
many-body problem given realistic nucleon-nucleon
interactions.
In no-core configuration interaction (NCCI)
calculations~\cite{navratil2000:12c-ncsm-COMBO,barrett2013:ncsm},
the nuclear many-body problem is formulated as a
matrix eigenproblem.  The Hamiltonian is represented in terms of basis states which
are antisymmetrized products of single-particle states for the full
$A$-body system of nucleons, \textit{i.e.}, with no assumption of an
inert core.  

In practice, the nuclear many-body calculation must be carried out in
a truncated space.  The dimension of the problem grows combinatorially
with the size of the included single-particle space and with the
number of nucleons in the system.  Computational restrictions
therefore limit the extent to which converged results can be obtained,
for energies or for other properties of the wave functions.  Except
for the very lightest systems ($A\lesssim 4$), convergence is
generally beyond reach.  Instead, we seek to approach convergence as
closely as possible.  Based on the still-unconverged calculations
which are computationally feasible, we would then ideally be able to
obtain a reliable estimate of the true values of observables which
would be obtained in the full, untruncated space.  Therefore, progress
may be pursued both by seeking accelerated convergence,
\textit{e.g.}, through the choice of basis, as considered here, and by
developing means by which robust extrapolations can be
made~\cite{maris2009:ncfc,coon2012:nscm-ho-regulator,furnstahl2012:ho-extrapolation,more2013:ir-extrapolation,jurgenson2013:ncsm-srg-pshell}.

NCCI calculations have so far been based almost
exclusively upon bases constructed from harmonic oscillator
single-particle wave functions.  The harmonic oscillator radial
functions have the significant limitation that their asymptotic
behavior is Gaussian, \textit{i.e.}, falling as $e^{-\alpha r^2}$ for
large $r$.  The actual asymptotics for nucleons bound by a
finite-range force are instead expected to be exponential,
\textit{i.e.}, falling as $e^{-\beta r}$.  
\begin{figure}
\centerline{\includegraphics[width=0.67\textwidth]{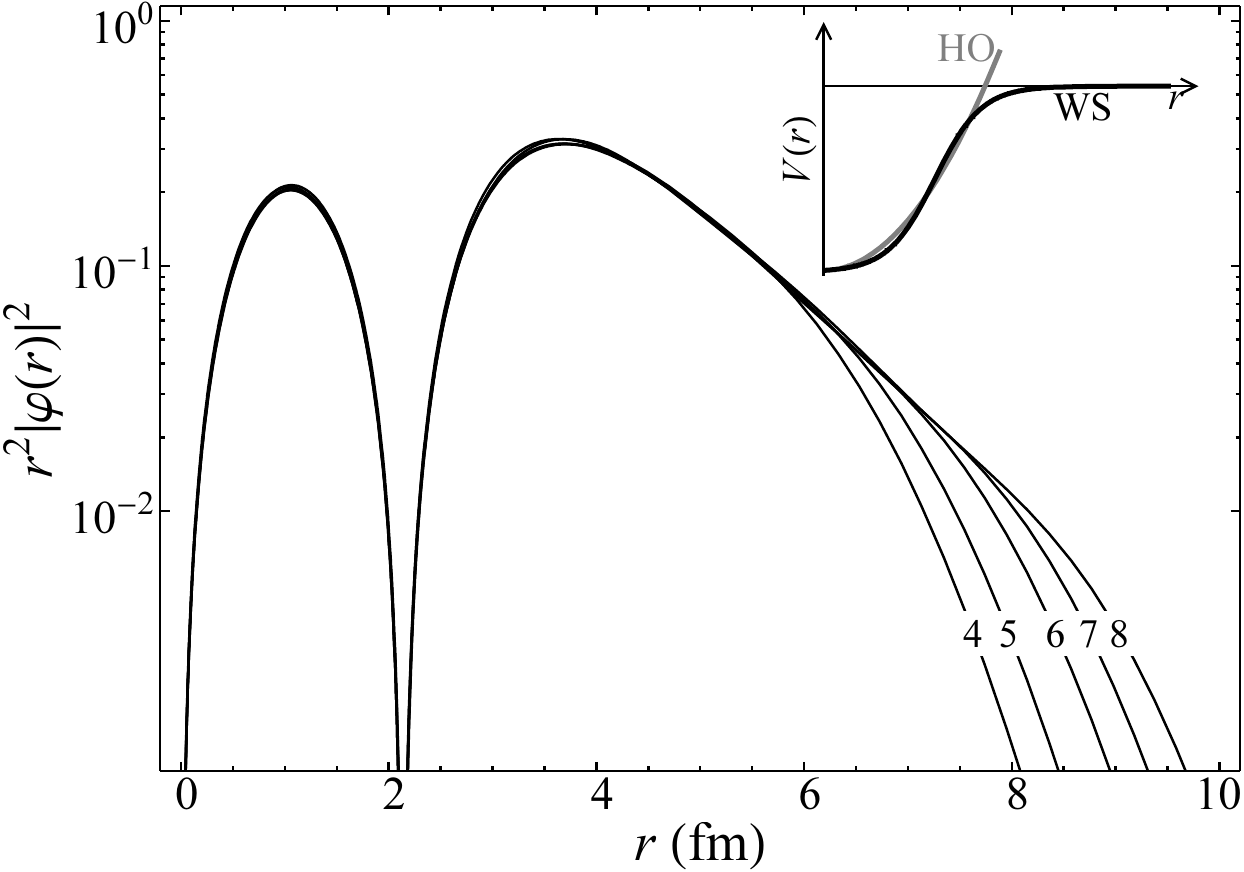}}
\caption{The calculated wavefunction obtained when a
problem with exponential asymptotics~--- here, the Woods-Saxon problem
is taken
for illustration~--- is solved in a finite basis of
oscillator functions.  The radial probability density $r^2\abs{\varphi(r)}^2$ is shown on a
logarithmic scale, so that exponential asymptotics would appear as a
straight line.  The
Woods-Saxon and oscillator potentials are shown in the inset.
(Solutions are for the Woods-Saxon $1s_{1/2}$  function, with  potential parameters appropriate to neutrons in
$\isotope[16]{O}$~\cite{suhonen2007:nucleons-nucleus}, with maximal
basis radial quantum numbers $n$ as
indicated.)  
}
\label{fig-ws-soln}      
\end{figure}

The problem encountered in using an oscillator basis to describe a
system with exponential asymptotics may be illustrated through the
simple one-dimensional example of the Schr\"odinger equation with a
Woods-Saxon potential.  In Fig.~\ref{fig-ws-soln}, we see the results
of solving for a particular eigenfunction in terms of successively
larger bases of oscillator radial functions.  In the classically
forbidden region, where the potential is nearly flat, the tail of the
wave function should be exponential.  It should thus appear as a
straight line on the logarithmic scale in Fig.~\ref{fig-ws-soln}.
Inclusion of each additional basis function yields a small extension
to the region in which the expected straight-line behavior is
reproduced, but, for any finite number of oscillator functions, there
is a radius beyond which the calculated tail is seen to sharply fall
below the true asymptotics.

Observables which are sensitive to the large-radius asymptotic
portions of the nuclear wave function therefore present a special
challenge to convergence in NCCI calculations with a conventional
oscillator basis.  Such ``long-range''
observables include the RMS radius and $E2$ moments and transitions, since the $r^2$
dependence of the relevant operators in both cases preferentially
weight the larger-$r$ portions of the wave-function.  The results for
these observables in NCCI calculations are in general highly
basis-dependent~\cite{bogner2008:ncsm-converg-2N,cockrell2012:li-ncfc}.

Furthermore, a prominent feature in light nuclei is the emergence of
halo structure~\cite{tanihata2013:halo-expt}, in
which one or more loosely-bound nucleons surround a compact core,
spending much of their time in the classically-forbidden region.  A
realistic treatment of the long-range properties of the
wave function is essential for an accurate reproduction of the halo
structure~\cite{quaglioni2009:ncsm-rgm}.

We are therefore motivated to consider alternative bases
which might be better suited for expanding the nuclear wave function
in its asymptotic region.  The framework for carrying
out NCCI calculations with a general radial basis is developed in
Ref.~\cite{caprio2012:csbasis}.  We explore the use of the Coulomb-Sturmian
functions~\cite{rotenberg1962:sturmian-scatt,weniger1985:fourier-plane-wave,keister1997:on-basis},
which form a complete set of square-integrable
functions and have exponential
asymptotics.  

In the present work, we apply the Colomb-Sturmian basis to NCCI
calculations for the neutron halo nuclei $\isotope[6,8]{He}$~--- as
well as to the baseline case $\isotope[4]{He}$, for which converged
results can be obtained.  We examine the possibility of extracting
RMS radii for the proton and neutron distributions based on a
relatively straightforward estimate, the ``crossover
point''~\cite{bogner2008:ncsm-converg-2N,cockrell2012:li-ncfc}, 
pending further development of more sophisticated
extrapolation schemes~\cite{furnstahl2012:ho-extrapolation}.  
Motivated by the disparity between proton and neutron radial
distributions in the neutron-rich halo nuclei, we also explore the use of
proton-neutron asymmetric bases, with different length scales for the proton and neutron radial basis
functions.  The basis and methods are first reviewed
(Sec.~\ref{sec-methods}), after which the results for
$\isotope[4,6,8]{He}$ are discussed (Sec.~\ref{sec-results}).

\section{Basis and methods}
\label{sec-methods}

\begin{figure}
\centerline{\includegraphics[width=\textwidth]{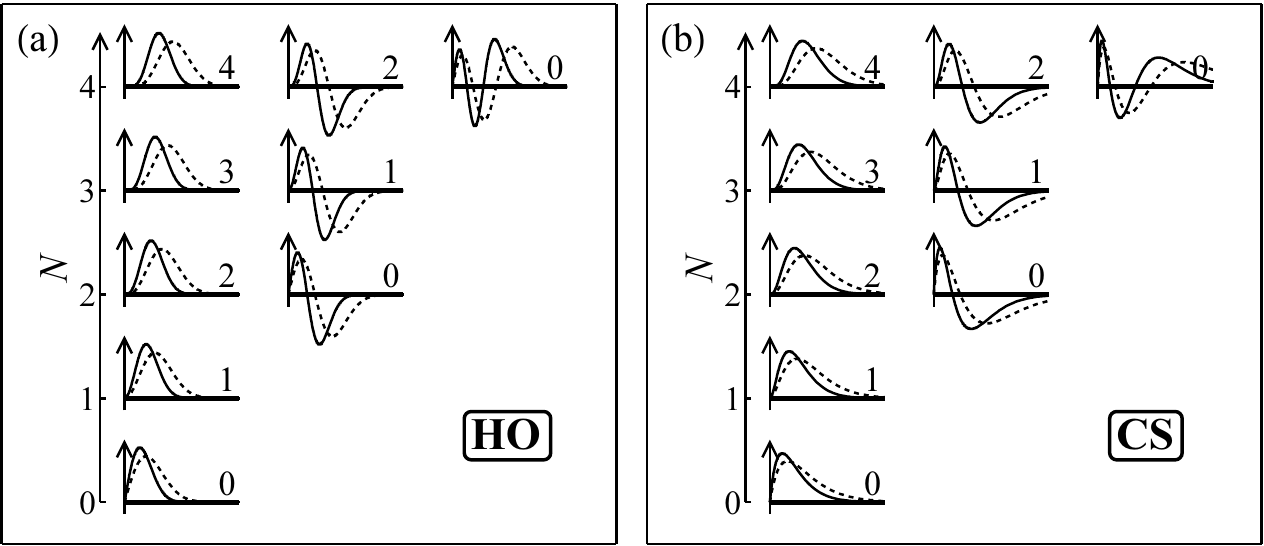}}
\caption{Radial functions (a)~$R_{nl}(b;r)$ of the harmonic oscillator
basis and (b)~$S_{nl}(b_l;r)$ of the Coulomb-Sturmian basis, with
$b_l$ given by the node-matching prescription (see text).  These
functions are shown arranged according to the harmonic oscillator
principal quantum number
$N\equiv 2n+l$, and are labeled by $l$.  The dotted curves
show the same functions dilated outward by a factor of
$\sqrt{2}\approx1.414$.
}
\label{fig-scheme-radial}      
\end{figure}

The harmonic oscillator basis functions, as used in conventional NCCI
calculations, constitute a complete, discrete, orthogonal set of
square-integrable functions and are given by
$\wfho_{nlm}(b;\mcvec{r})=R_{nl}(b;r)Y_{lm}(\uvec{r})/r$, with radial
wave functions 
\begin{equation}
\label{eqn-ho-R}
R_{nl}(b;r)\propto(r/b)^{l+1}
L_n^{l+1/2}[(r/b)^2]
e^{-\tfrac12(r/b)^2},
\end{equation}
where the $L_n^\alpha$ are generalized Laguerre polynomials, the
$Y_{lm}$ are spherical harmonics, $n$ is the radial quantum number,
$l$ and $m$ are the orbital angular momentum and its $z$-projection, and
$b$ is the oscillator length.  
The Coulomb-Sturmian
functions
likewise constitute a complete, discrete, orthogonal set of square-integrable
functions, while also possessing exponential asymptotics
more appropriate to the nuclear problem.  
They are given by
$\Lambda_{nlm}(b;\mcvec{r})=S_{nl}(b;r)Y_{lm}(\uvec{r})/r$,
with radial wave functions 
\begin{equation}
\label{eqn-cs-S}
S_{nl}(b;r)\propto(2r/b)^{l+1}
L_n^{2l+2}(2r/b)
e^{-r/b},
\end{equation}
where $b$ again represents a length scale.  Further details may be found
in Ref.~\cite{caprio2012:csbasis}.  
Both sets of radial functions are shown in Fig.~\ref{fig-scheme-radial}, for comparison.

For either basis, the single-particle basis states $\tket{nljm}$ are
then defined by coupling of the orbital angular momentum with the
spin, to give total angular momentum $j$, and the many-body basis is
defined by taking antisymmetrized products of these single-particle
states.  Thus, the structure of the many-body calculation is
independent of the details of the radial basis.  The choice of
radial basis only enters the calculation through the values of the
Hamiltonian two-body matrix elements (or higher-body matrix elements,
if present), which we must first generate as the input to the
many-body calculation.

The nuclear Hamiltonian for NCCI calculations has the form $H=\Trel +
V$, where $\Trel$ is the Galilean-invariant, two-body relative kinetic
energy operator, and $V$ is the nucleon-nucleon
interaction.\footnote{A Lawson term proportional the number $\Ncm$ of
center-of-mass oscillator quanta can also be included, to shift
center-of-mass excitations out of the low-lying spectrum, but it is
not essential for the ground-state properties considered here.  The
implications of center-of-mass dynamics for general bases are
addressed in Ref.~\cite{caprio2012:csbasis}.}  The relative kinetic
energy decomposes into one-body and two-body terms as
\begin{equation}
\label{eqn-Trel}
\Trel\equiv\frac{1}{4Am_N} \sumprime_{ij} (\mcvec{p}_i-\mcvec{p}_j)^2\\
=\frac{1}{2Am_N}\biggl[(A-1) 
\sum_i \mcvec{p}_i^2
- \sumprime_{ij} \mcvec{p}_i\cdot\mcvec{p}_j
\biggr].
\end{equation}
Since the two-body term is separable, 
matrix elements of $\Trel$ may be calculated in a straightforward fashion for any radial basis, in
terms of one-dimensional radial integrals of the operators $p$ and $p^2$~\cite{caprio2012:csbasis}.

Calculation of the interaction two-body matrix elements becomes more
involved if one moves to a general radial basis.  The nucleon-nucleon
interaction is defined in relative coordinates.  The oscillator basis
is special, in that matrix elements in a relative oscillator basis, consisting of
functions $\wfho_{nl}(\mcvec{r}_1-\mcvec{r}_2)$, can readily be
transformed to the two-body oscillator basis, consisting of functions
$\wfho_{n_1l_1}(\mcvec{r}_1)\wfho_{n_2l_2}(\mcvec{r}_2)$, by the Moshinsky
transformation.  We therefore still begin by carrying out the transformation to two-body matrix elements
$\tme{cd;J}{V}{ab;J}$ with respect to the oscillator basis, and only then carry
out a change of basis to the Coulomb-Sturmian basis in the two-body space, as~\cite{caprio2012:csbasis}
\begin{equation}
\label{eqn-tbme-xform}
\tme{\bar{c}\bar{d};J}{V}{\bar{a}\bar{b};J}=
\sum_{abcd} \toverlap{a}{\bar{a}}\toverlap{b}{\bar{b}}
\toverlap{c}{\bar{c}}\toverlap{d}{\bar{d}}
\, \tme{cd;J}{V}{ab;J},
\end{equation}
where we label single-particle orbitals for the oscillator basis by
unbarred symbols $a=(n_al_aj_a)$ and those for the Coulomb-Sturmian
basis by barred symbols $\bar{a}=(\bar n_a
\bar l_a \bar j_a)$.
The coefficients $\toverlap{a}{\bar{a}}$ required for the
transformation are obtained from straightforward one-dimensional overlaps of the harmonic
oscillator and Coulomb-Sturmian radial functions,
$\toverlap{R_{nl}}{S_{\bar{n}l}} = \int_0^\infty dr\, R_{nl}(\bint;r)
S_{\bar{n}l}(b;r)$.  The oscillator length $\bint$ with respect to
which the interaction two-body matrix elements are defined and the
length scale $b$ of the final Coulomb-Sturmian basis functions may in
general be different.  The change-of-basis transformation
in~(\ref{eqn-tbme-xform}) is, in practice, limited to a finite sum,
\textit{e.g.}, with a shell cutoff $N_a,N_b,N_c,N_d\leq\Ncut$. The
cutoff $\Ncut$ must be chosen high enough to insure that the results
of the subsequent many-body calculation are cutoff-independent, which
may in general depend upon the oscillator and Coulomb-Sturmian length
parameters, interaction, nucleus, and observable at hand.

Any single particle basis,
including~(\ref{eqn-ho-R}) or~(\ref{eqn-cs-S}), has a free length
scale $b$.  For the oscillator basis, this is traditionally quoted as
the oscillator energy $\hbar\Omega$, where 
\begin{equation}
\label{eqn-beta-bho}
b(\hbar\Omega)=\frac{(\hbar c)}{[(m_Nc^2)(\hbar\Omega)]^{1/2}}.
\end{equation}
In deference to the convention of presenting NCCI results as a
function of the basis ``$\hbar\Omega$'', we nominally carry over this
relation to define an $\hw$ parameter for general radial bases, although $\hbar\Omega$ no longer has
any direct physical meaning as an energy scale.  Regardless, the
inverse square-root dependence remains, so that a factor of two change in $\hbar\Omega$
describes a factor of $\sqrt{2}$ change in radial scale, as
illustrated for both harmonic oscillator and Coulomb-Sturmian
bases by the dotted curves in Fig.~\ref{fig-scheme-radial}.  

Furthermore, there is much additional freedom in the basis, since the many-body basis states (antisymmetrized product
states) constructed from a single-particle basis are orthonormal so
long as the single-particle states are orthonormal.  Orthogonality for
single-particle states of different $l$ or $j$ follows entirely from
the angular and spin parts of the wave function.  Only orthogonality
\textit{within} the space of a given $l$ and $j$ follows from the
radial functions, \textit{e.g.}, for the Coulomb-Sturmian functions,
$\toverlap{n'l'j'}{nlj}=\bigl[\int
dr\,S_{n'l}(b;r)\,S_{nl}(b;r)\bigr]\,\delta_{l'l}\delta_{j'j}$.  We
are therefore free to choose $b$ independently, firstly, for each $l$
space (or $j$ space), as $b_l$ (or $b_{lj}$), and, secondly, for
protons and neutrons, as $b_p$ and $b_n$.

The first observation raises the possibility, still to be explored, of
obtaining significant improvements in the efficacy of the basis by
optimizing the $l$-dependence of the length parameter.  In Ref.~\cite{caprio2012:csbasis}, the radial scale of the
Coulomb-Sturmian functions, for each $l$, was fixed by matching the first node of
the $n=1$ Coulomb-Sturmian function  to the first node of
the $n=1$  oscillator
function, at that $l$, yielding the prescription $b_l=[2/(2l+3)]^{1/2}b(\hw)$~\cite{caprio2012:csbasis}.

The second observation raises the possibility of proton-neutron
asymmetric length scales, which might be advantageous for nuclei
with significant disparities between the proton and neutron
distributions, in particular, halo nuclei.
Therefore, in the present work, we adopt
\begin{equation}
\label{eqn-blpn}
{b_{l,p}}=\sqrt{\frac{2}{2l+3}}{b}({{\hbar\Omega}}) \qquad {b_{l,n}}={\beta}\sqrt{\frac{2}{2l+3}}{b}({{\hbar\Omega}}),
\end{equation}
where $\beta$ sets an overall relative scale $b_n/b_p$.  For
example, if the solid and dotted curves in
Fig.~\ref{fig-scheme-radial}(b) are taken to represent the proton and
neutron radial functions, respectively, then the figure illustrates
the case in which $b_n/b_p=\sqrt{2}\approx1.414$.


\section{Results for the \boldmath$\isotope{He}$ isotopes}
\label{sec-results}

\begin{figure}[p]
\centerline{\includegraphics[width=0.9\textwidth]{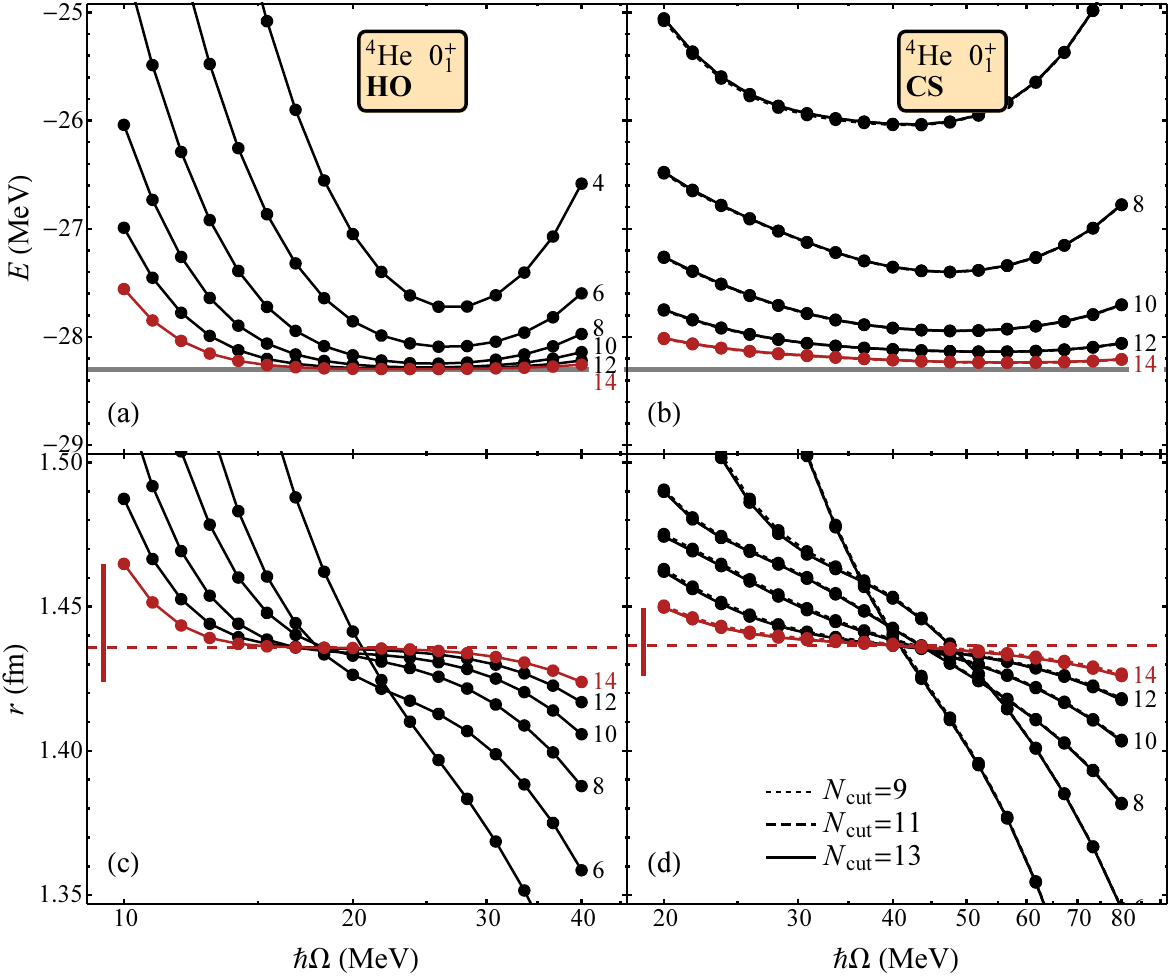}}
\caption{The calculated $\isotope[4]{He}$ ground state energy~(top)
and RMS point-proton radius $r_p$~(bottom), using the
conventional oscillator~(left) and Coulomb-Sturmian~(right) bases.
These are shown as functions of the
basis $\hbar\Omega$ parameter, for $\Nmax=4$ to $14$ (as labeled),
and for
transformation cutoffs $\Ncut=9$, $11$, and $13$ (Coulomb-Sturmian basis only, indicated by dashing, curves
nearly indistinguishable).  The
converged energy is indicated by the horizontal line (at top), 
the crossover radii by  dashed horizontal lines (at bottom), and the 
spread in radius values by vertical bars (again at bottom).
}
\label{fig-4he-scan}      
\end{figure}

\begin{figure}[p]
\centerline{\includegraphics[width=0.7\textwidth]{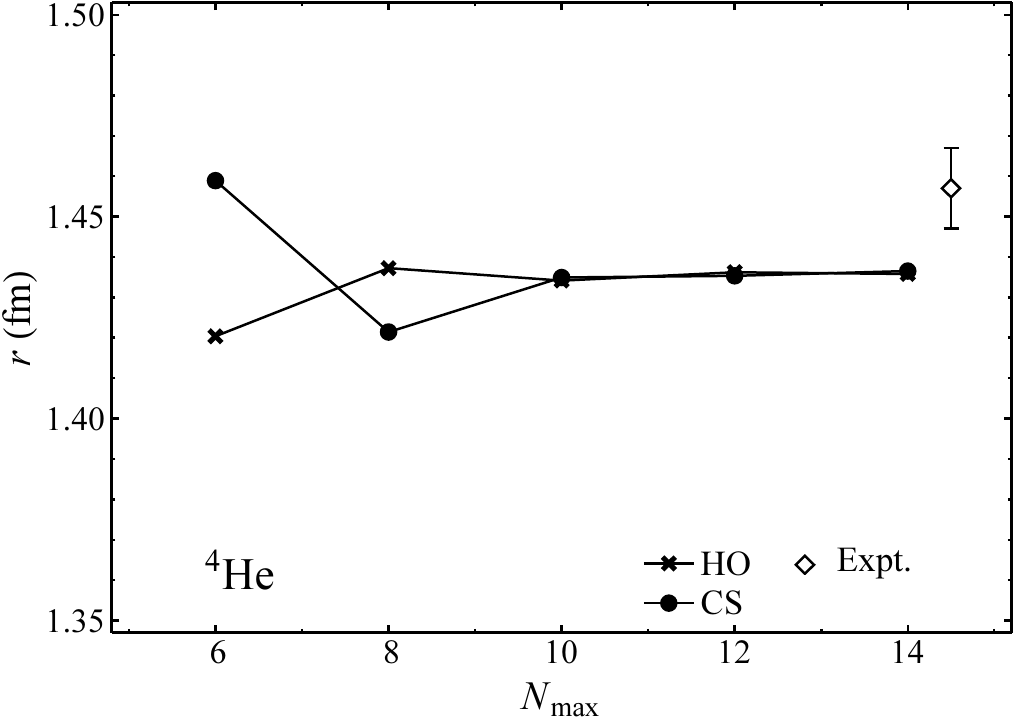}}
\caption{The $\isotope[4]{He}$ ground state RMS point-proton radius $r_p$,
as estimated from the crossover point (see text), calculated for
the harmonic oscillator and Coulomb-Sturmian bases.  The experimental
value is from Ref.~\cite{tanihata2013:halo-expt}.
}
\label{fig-4he-nmax}      
\end{figure}

We carry out calculations for the isotopes $\isotope[4,6,8]{He}$ using both the harmonic
oscillator and Coulomb-Sturmian bases.
These calculations are based on the JISP16 nucleon-nucleon
interaction~\cite{shirokov2007:nn-jisp16}, plus Coulomb
interaction. The bare interaction is used, \textit{i.e.}, without renormalization.  The
proton-neutron $M$-scheme code
MFDn~\cite{maris2010:ncsm-mfdn-iccs10,aktulga2013:mfdn-ONLINE}
is employed for the many-body calculations.  Results are calculated
with basis truncations up to $\Nmax=14$ for $\isotope[4]{He}$, $\Nmax=12$ for $\isotope[6]{He}$,
and $\Nmax=10$ for $\isotope[8]{He}$.\footnote{The harmonic oscillator many-body basis is normally truncated according to the $\Nmax$ scheme, based on the total number of oscillator quanta.  
That is, the many-body basis states are characterized by a total
number of oscillator quanta $\Ntot\equiv\sum_i N_i$, where
$N_i\equiv 2n_i+l_i$.  If $\Ntot$ is written as $\Ntot = N_0+\Nex$, where $N_0$ in the lowest Pauli-allowed number of
quanta, then the basis is subject to the restriction $\Nex\leq\Nmax$.
We formally
carry this truncation over to the Coulomb-Sturmian basis, although
$N\equiv 2n+l$ no longer has significance as an oscillator principal quantum
number.}

The last neutrons in $\isotope[6]{He}$ and $\isotope[8]{He}$ are only
weakly bound, with two-neutron separation energies of $0.97\,\MeV$ and
$2.14\,\MeV$, respectively.  These isotopes are interpreted as
consisting of neutron halos surrounding an $\alpha$
core~\cite{tanihata2013:halo-expt}.  The basic observables indicating
halo properties are the RMS radii of the proton and neutron
distributions, $r_p$ and $r_n$, respectively.\footnote{Specifically,
$r_p$ and $r_n$ are the RMS radii of the point-proton and
point-neutron distributions, measured relative to the center of mass.
See Ref.~\cite{bacca2012:6he-hyperspherical} for definitions, and
Ref.~\cite{caprio2012:csbasis} for evaluation of the two-body relative
RMS radius observable with a general radial basis.  From the analysis
of experimental charge and matter radii in
Ref.~\cite{tanihata2013:halo-expt}, $\isotope[4]{He}$ has
$r_p=1.457(10)\,\fm \,(\approx r_n)$, $\isotope[6]{He}$ has
$r_p=1.925(12)\,\fm$ and $r_n=2.74(7)\,\fm$, and $\isotope[8]{He}$ has
$r_p=1.807(28)\,\fm$ and $r_n=2.72(4)\,\fm$.}  Moving from
$\isotope[4]{He}$ to $\isotope[6]{He}$, $r_p$ increases by $\sim32\%$.
This may be understood as resulting from the recoil of the $\alpha$
core against the halo neutrons, and potentially core polarization, as
well.  In turn, $r_n$ is larger than $r_p$ by $\sim42\%$, reflecting
the extended halo neutron distribution.  The radii for
$\isotope[8]{He}$ are comparable to those for $\isotope[6]{He}$.

We first consider calculations for 
$\isotope[4]{He}$ as a baseline.  Results are shown over two doublings in $\hw$,
\textit{i.e.}, representing a doubling in basis length scale, in
Fig.~\ref{fig-4he-scan}.  Energy convergence is reached for the
harmonic oscillator basis, as evidenced by approximate $\Nmax$ and
$\hw$ independence of the higher $\Nmax$ results over a range of
$\hw$ values, in Fig.~\ref{fig-4he-scan}(a,b).  Convergence is obtained at the
$\sim10\,\mathrm{keV}$ level by $\Nmax=14$.  The binding energies
for $\isotope[4]{He}$ computed with the Coulomb-Sturmian basis lag significantly behind
those obtained with the oscillator basis, by about two steps in
$\Nmax$.  This should perhaps not be surprising, given that
$\isotope[4]{He}$ is tightly bound, and the structure can thus be
expected to be driven
by short-range correlations rather than asymptotic properties.
Incidentally, it may be seen from Fig.~\ref{fig-4he-scan}(b,d) that
stability with respect to the cutoff in the change-of-basis
transformation~(\ref{eqn-tbme-xform}) has been obtained~---
calculations with $\Ncut=9$, $11$, and $13$ are
virtually indistinguishable (the transformation has been carried
out from
oscillator basis interaction matrix elements at $\hwint=40\,\MeV$).

Convergence of the computed RMS radii, for both the oscillator and
Coulomb-Sturmian bases, is again indicated by approximate $\Nmax$ and
$\hw$ independence over a range of $\hw$ values, which appears as a
shoulder in the curves of Fig.~\ref{fig-4he-scan}(c,d).  The $\hw$
dependence for the Coulomb-Sturmian calculations appears to be moderately
shallower, over the range (two doublings) of $\hw$ shown, than for the
harmonic oscillator calculations [see vertical bars
in Fig.~\ref{fig-4he-scan}(c,d)].  

It was proposed in
Refs.~\cite{bogner2008:ncsm-converg-2N,cockrell2012:li-ncfc} that the
radius can be estimated~--- even
before convergence is well-developed~--- by the crossover point
between the curves
obtained for successive $\Nmax$ values.  This is an admittedly
\textit{ad hoc} prescription, rather than a theoretically motivated
extrapolation.  However, we can test it~--- for both oscillator and
Coulomb-Sturmian bases~--- in this case of $\isotope[4]{He}$, where the
final converged value is known.  The crossover radii are shown as a
function of $\Nmax$, for both bases, in
Fig.~\ref{fig-4he-nmax}.  The curves used in deducing these crossovers are computed by cubic
interpolation of the calculated data points at different $\hw$.  The crossovers
already serve to estimate the final converged value to within $\sim0.05\,\fm$
at $\Nmax=6$.
It may be noted, from Fig.~\ref{fig-4he-nmax}, that the converged
radius obtained with the JISP16 interaction agrees with experiment to
within $\sim0.03\,\fm$.

\begin{figure}[p]
\centerline{\includegraphics[width=1.0\textwidth]{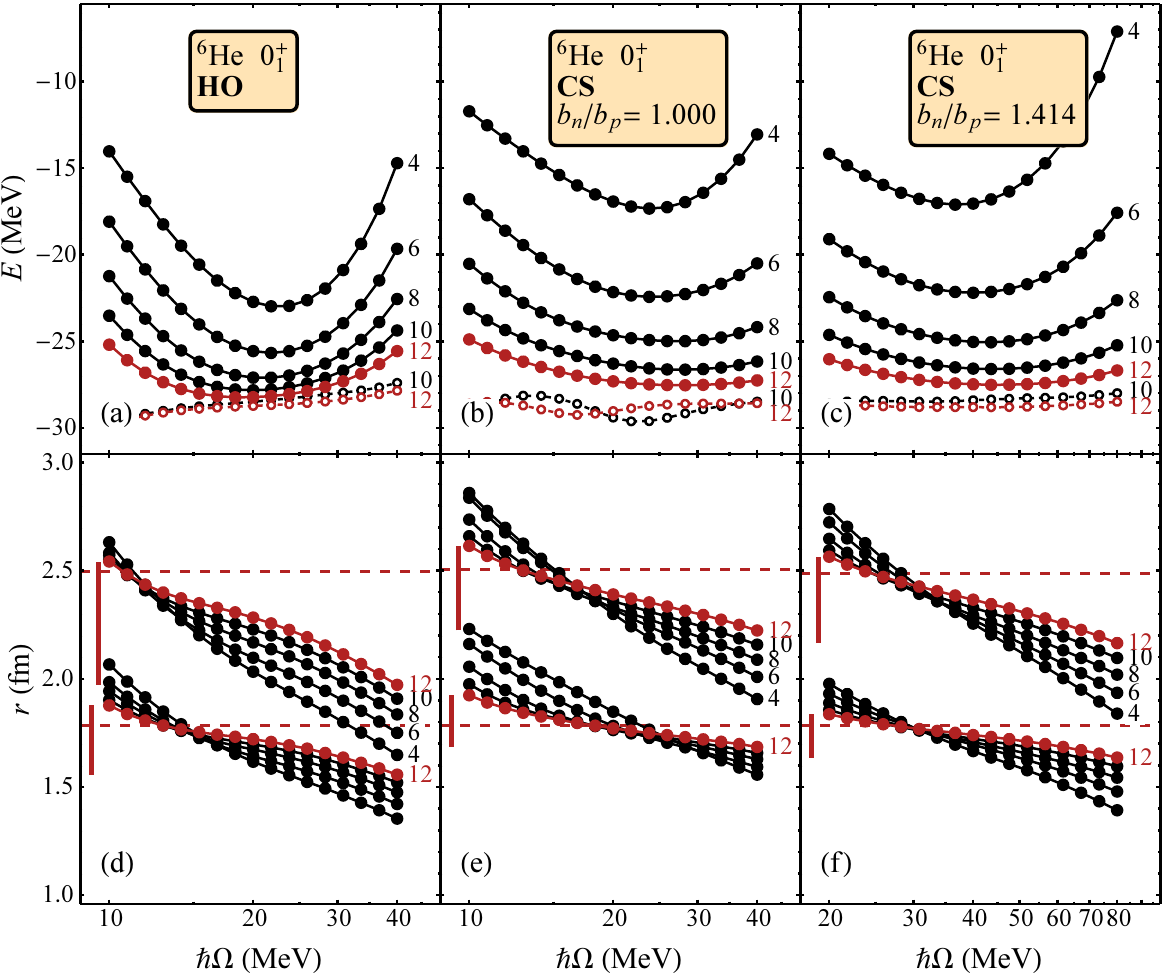}}
\caption{The calculated $\isotope[6]{He}$ ground state energy~(top)
and RMS point-proton radius $r_p$ and point-neutron radius $r_n$~(bottom), using the
conventional oscillator basis~(left), Coulomb-Sturmian
basis~(center), and Coulomb-Sturmian
basis with $b_n/b_p=1.414$~(right).   Exponentially extrapolated energies
are indicated by dashed curves (at top), and crossover radii by dashed
horizontal lines (at bottom).
}
\label{fig-6he-scan}      
\end{figure}

\begin{figure}[p]
\centerline{\includegraphics[width=1.0\textwidth]{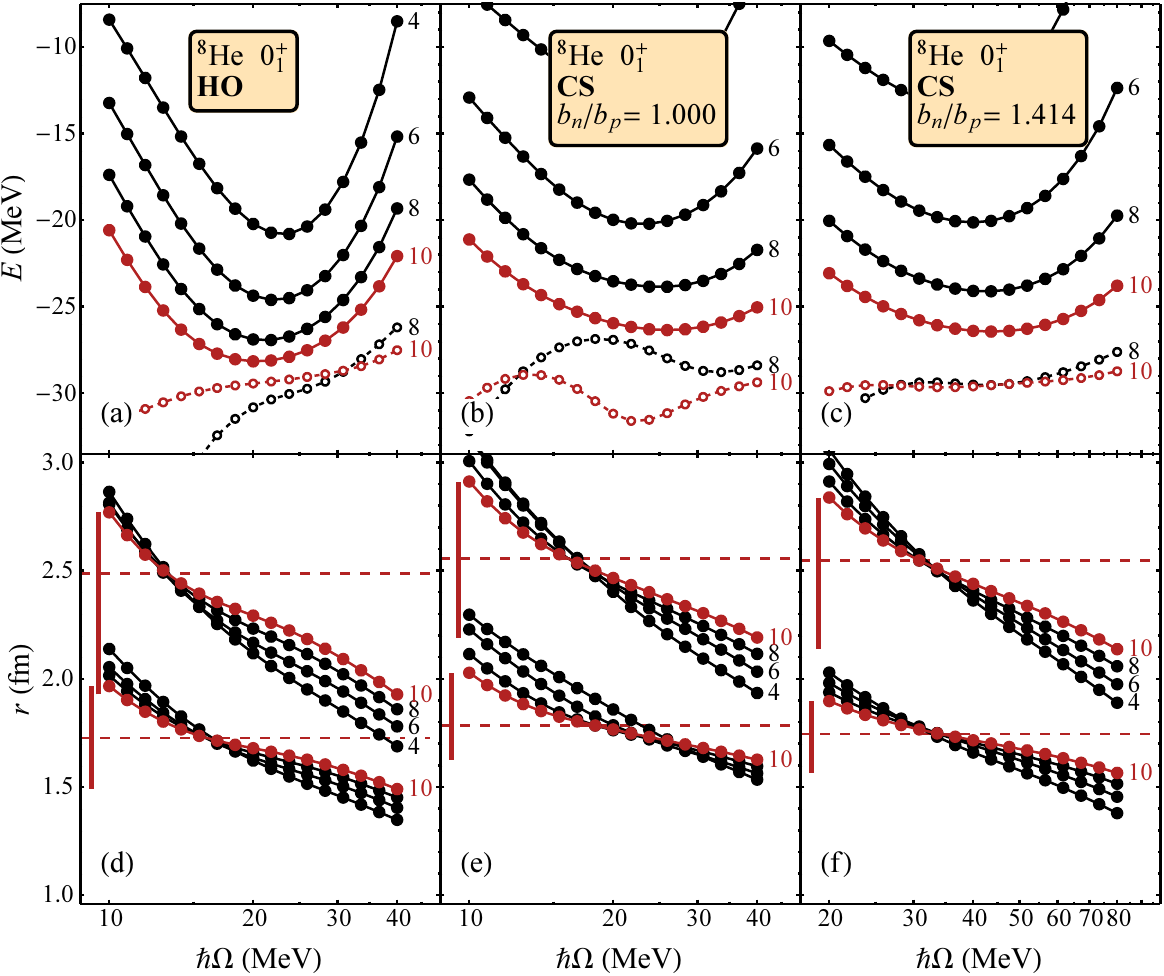}}
\caption{The calculated $\isotope[8]{He}$ ground state energy~(top)
and RMS point-proton radius $r_p$ and point-neutron radius $r_n$~(bottom), using the
conventional oscillator basis~(left), Coulomb-Sturmian
basis~(center), and Coulomb-Sturmian
basis with $b_n/b_p=1.414$~(right).  Exponentially extrapolated energies
are indicated by dashed curves (at top), and crossover radii by dashed
horizontal lines (at bottom).
}
\label{fig-8he-scan}      
\end{figure}

Let us now consider the calculations for the halo nuclei $\isotope[6,8]{He}$.  The computed
ground state energies, proton radii, and neutron radii are shown in
Figs.~\ref{fig-6he-scan} and~\ref{fig-8he-scan}.  Results are included
(at right in each figure) for a
Coulomb-Sturmian basis with proton-neutron asymmetric length scales in
the ratio $b_n/b_p=1.414$, which is comparable to
the ratio $r_n/r_p$ of neutron and proton distribution radii for these
nuclei.  Energy convergence in the Coulomb-Sturmian basis lags that of
the harmonic oscillator basis, but less dramatically than seen above for
$\isotope[4]{He}$.  A basic three-point exponential extrapolation of
the energy with respect to $\Nmax$, at each $\hw$ value, is indicated
by the dashed curves in Figs.~\ref{fig-6he-scan}
and~\ref{fig-8he-scan}.  The extrapolated energy is remarkably
$\hw$-independent in the $b_n/b_p=1.414$ calculations, still with some
$\Nmax$ dependence.  It appears to be approximately consistent with the harmonic
oscillator extrapolations as well.  However, such extrapolations must be
viewed with caution, as both theoretical arguments and empirical studies
suggest that other functional forms may be more appropriate, over at least
portions of the $\hw$
range~\cite{coon2012:nscm-ho-regulator,furnstahl2012:ho-extrapolation,more2013:ir-extrapolation}.

\begin{figure}[p]
\centerline{\includegraphics[width=0.7\textwidth]{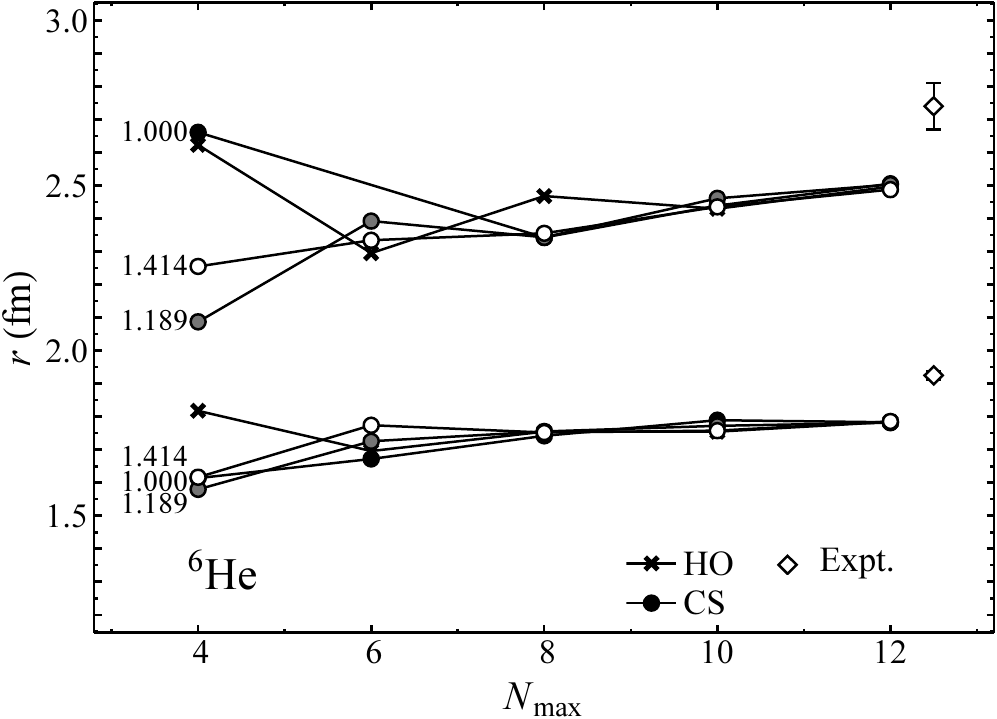}}
\caption{The $\isotope[6]{He}$ ground state RMS point-proton radius
$r_p$ (lower
curves) and point-neutron radius $r_n$ (upper curves),
as estimated from the crossover point (see text), calculated for
the harmonic oscillator basis and for Coulomb-Sturmian bases with
$b_n/b_p=1$, $1.189$, and $1.414$ (as indicated).  Experimental values are from Ref.~\cite{tanihata2013:halo-expt}.
}
\label{fig-6he-nmax}      
\end{figure}

\begin{figure}[p]
\centerline{\includegraphics[width=0.7\textwidth]{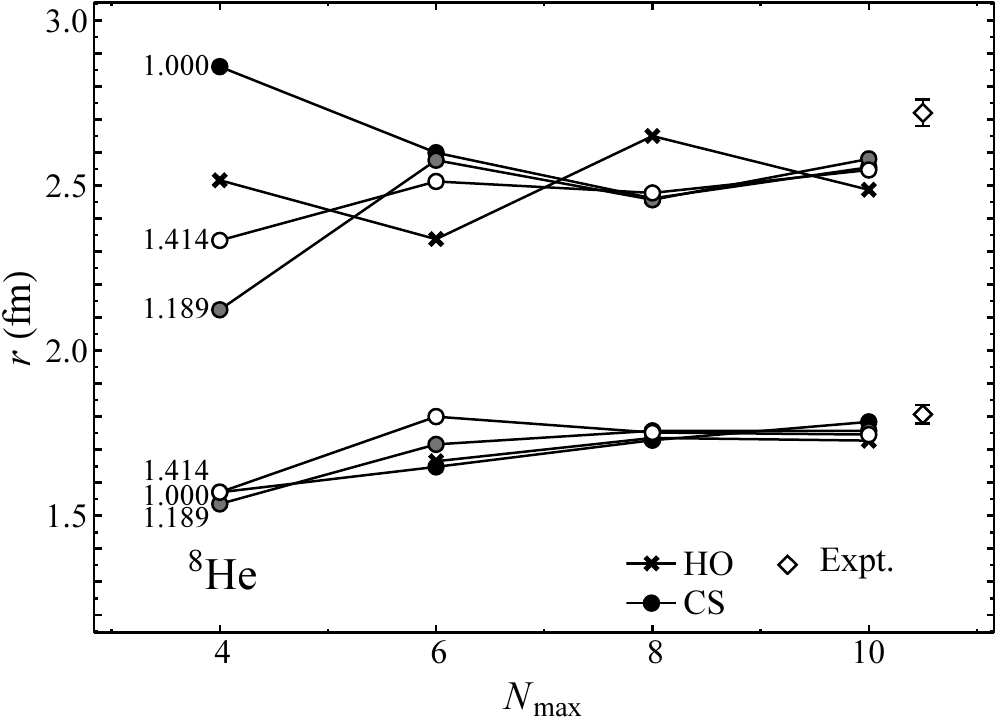}}
\caption{The $\isotope[8]{He}$ ground state RMS point-proton radius
$r_p$ (lower
curves) and point-neutron radius $r_n$ (upper curves),
as estimated from the crossover point (see text), calculated for
the harmonic oscillator basis and for Coulomb-Sturmian bases with
$b_n/b_p=1$, $1.189$, and $1.414$ (as indicated).  Experimental values are from Ref.~\cite{tanihata2013:halo-expt}.
}
\label{fig-8he-nmax}      
\end{figure}

Comparing the results for radii obtained 
with the various bases, for $\isotope[6,8]{He}$, we see that the
Coulomb-Sturmian results (for either $b_n/b_p=1$ or $b_n/b_p=1.414$)
again have a moderately shallower $\hw$ dependence than obtained with
the harmonic oscillator basis.  Well-defined and stable crossover
points are visible in Figs.~\ref{fig-6he-scan} and~\ref{fig-8he-scan},
especially for the $b_n/b_p=1.414$ calculations (at right).  The
extracted crossover radii are shown, as functions of $\Nmax$, in
Figs.~\ref{fig-6he-nmax} and~\ref{fig-8he-nmax}.  The radii obtained
for the Coulomb-Sturmian calculations with different ratios of neutron
and proton length scales ($b_n/b_p=1$, $1.189$, and $1.414$) track
each other closely from $\Nmax\approx8$ onward, agreeing with each
other to within $\sim0.1\,\fm$.  For $r_p$, the values are stable with respect to $\Nmax$ and agree
with the values obtained from the harmonic oscillator basis crossover
as well.  For $r_n$, it appears that the values might be
drifting systematically with $\Nmax$, although they do remain within an
$\sim0.2\,\fm$ range from $\Nmax=6$ onward.  (The crossover radii obtained from the harmonic oscillator calculations are
 fluctuating over a wider range.)  Therefore, it is not possible to give a
definitive value, but an estimate of $r_n\approx2.5\text{--}2.6\,\fm$
can reasonably be made, for both $\isotope[6,8]{He}$.  

Thus, \textit{ab initio} NCCI calculations for $\isotope[6,8]{He}$
with the JISP16 interaction, using both conventional and
Coulomb-Sturmian bases, yield consistent estimates of the RMS
point-proton and point-neutron radii, when these are extracted by the
crossover prescription.  The results qualitatively reproduce the trend
in proton and neutron radii across the $\isotope{He}$ isotopes, while
quantitatively suggesting that the JISP16 interaction may yield radii
which are smaller than experimentally observed, by as much as
$\sim0.2\text{--}0.3\,\fm$ for the $\isotope[6,8]{He}$ neutron radii.

\clearpage
\section*{Acknowledgements}
\sloppy 

We thank S.~Quaglioni, S.~Bacca, and M.~Brodeur for
valuable discussions and A.~E.~McCoy for comments on the manuscript.  This work was supported by the Research
Corporation for Science Advancement 
through the 
Cottrell Scholar
program, by the US Department of Energy under Grants
No.~DE-FG02-95ER-40934, DESC0008485
(SciDAC/NUCLEI), and
DE-FG02-87ER40371, and by the US National Science Foundation under
Grant No.~0904782. Computational resources were provided by the
National Energy Research Supercomputer Center (NERSC), 
which is
supported by the Office of Science of the U.S. Department of Energy
under Contract No.~DE-AC02-05CH11231.


\providecommand{\APSLONG}{}
\providecommand{\ELSEVIER}{}


\end{document}